\documentstyle[preprint,aps,amsfonts]{revtex}

\draft

\def\cala{{\cal A}}
\def\vcala{\bbox{\cal A}}
\def\vcalp{\bbox{\cal P}}
\def\vhcalp{\bbox{\hat {\cal P}}}
\def\vcalsa{\bbox{\frak a}}

\def\cale{{\cal E}}
\def\vcale{\bbox{\cal E}}
\def\vec#1{{\bf#1}} \def\hatn#1{\hat{\bf#1}} \def\TT#1{\vec T_{#1}}
\def\vO{\vec \Omega} \def\den{n_{e}}

\begin{document}





\title{Response Function of the  Fractional Quantized Hall State
  on a Sphere I : Fermion Chern-Simons Theory} \author{Steven H. Simon and
Bertrand I. Halperin}

\bigskip \date{\today} \address{ Physics Department, Harvard
  University, Cambridge MA 02138 }

\maketitle \bigskip

\begin{abstract}

  Using a well known singular gauge transformation, certain fractional
  quantized Hall states can be modeled as integer quantized Hall
  states of transformed fermions interacting with a Chern-Simons
  field.  In previous work we have calculated the
  electromagnetic response function of these states at arbitrary
  frequency and wavevector by using the Random Phase Approximation
  (RPA) in combination with a Landau Fermi Liquid approach.  We now
  adopt these calculations to a spherical geometry in order to
  facilitate comparison with exact diagonalizations performed on
  finite size systems.

\end{abstract}
\newpage


\section{Introduction}

Calculational methods based on transforming electrons to fermions
interacting with a Chern-Simons field have shown themselves to be very
powerful for understanding fractional quantized Hall states and
related unquantized states\cite{Simon,Tome,Lopez}.  In a previous
paper\cite{Simon} we have developed a set of approximations for
calculating the current and density response functions within this
fermion Chern-Simons model at filling fractions corresponding to the
principal fractional quantized Hall states.  In particular, we use the
Random Phase Approximation (RPA) to account for fluctuations in the
Chern-Simons field, as well as to account for the direct Coulomb
interaction.  The large effective mass renormalization that we expect
to occur in the Chern-Simons theory is then accounted for by using a
Landau Fermi liquid theory approach.  This set of approximations,
which we call the ``Modified RPA,'' has a number of desirable
features.  To begin with, the $f$-sum rule, a result of Galilean
invariance, is automatically satisfied.  In addition, Kohn's theorem
--- which says that a mode at the bare cyclotron frequency must have
all the weight of the $f$-sum rule in the long wavelength limit --- is
also satisfied within this approach.  Finally, this approach predicts
a discrete series of quasiexciton lines, with the lowest branch in
qualitative agreement with exact diagonalization of finite size
systems\cite{Simon}.  More quantitative comparison of this theory with
the exact numerical work could not be made since the theory considers
an infinite planar system, whereas exact diagonalizations necessarily
consider only small systems.  The purpose of the present work is to
formulate an analogous fermion Chern-Simons theory on a sphere, so as
to facilitate comparison with exact diagonalization results.  A
quantitative comparison between our results and the results of exact
diagonalization of small spherical systems will be made in a following
paper\cite{Song}.

The outline of this paper is as follows.  In Sec. \ref{sub:mono} we
begin by discussing the Dirac string and the Dirac
quantization condition for a spherical system with a magnetic
monopole in its center.  We then write down the Hamiltonian that
describes interacting electrons on the surface of this sphere in
Sec. \ref{sub:Define}.  In Sec. \ref{sub:CST} a singular gauge
transformation is made that maps our system into a system of
``composite fermions'' --- fermions bound to an even number of
Chern-Simons flux quanta.  We treat the Chern-Simons flux within mean
field theory in Sec. \ref{sub:MFT} and  review how certain fractional
quantized Hall states can be viewed as integer quantized Hall states of
these composite fermions as originally proposed by Jain\cite{Jain}.

In Secs.  \ref{sub:Basis} and \ref{sub:DefofK} we define the current
and density response function in terms of a convenient spherical
basis.  We make use here of some restrictions on the form of the
response function which are derived in Appendices \ref{app:rot} and
\ref{app:cons} using rotational symmetry, gauge invariance, and
current conservation.  In Sec.  \ref{sub:MFR} we use standard linear
response theory to calculate the response function within the mean
field approximation.  This mean field response also serves as a
starting point for the RPA calculation.  (The most tedious part of the
mean field calculation, the evaluation of a correlation function, is
relegated to Appendices \ref{app:ret} and \ref{app:ident}.)  In Sec.
\ref{sub:SCRPA} we use a self consistent approximation to derive the
RPA equation for the response function.  (Certain angular momentum
components of the Coulomb and Chern-Simons interactions are evaluated
in Appendix \ref{app:VW}, and are used to establish the form of the
induced effective electric and magnetic fields in Appendix
\ref{app:EB}.)  In Sec.  \ref{sub:Mass} we discuss the issue of
quasiparticle effective mass renormalization.  In general, we expect
that the RPA results will either not properly account for this mass
renormalization, or will violate the $f$-sum rule.  To repair this
problem, we follow the results of Ref.  \cite{Simon} and give a
Modified RPA prescription that incorporates the effects of mass
renormalization and satisfies the $f$-sum rule.  Finally, in Sec.
\ref{sec:summary} we give a very brief summary of our work.

\section{The Basic Problem}

\label{sec:Basic}

\subsection{The Magnetic Monopole}
\label{sub:mono}
We begin by considering some of the features of a spherical system
with a magnetic monopole in its center.  Since we want the magnetic
field to be given by the curl of a vector potential, and the
divergence of a curl is zero, a monopole must correspond to a vector
potential that has a singularity.  Specifically, a one dimensional
``Dirac string,'' which is essentially an infinitely thin flux tube
carrying magnetic flux corresponding to the monopole's magnetic charge
{\it into} the monopole, stretches from the monopole to spatial
infinity\cite{Dirac}. The singular string can be moved, but not
eliminated, by a gauge transformation.  Formally, one should consider
the vector potential of the monopole in two different gauges
simultaneously such that the vector potentials is well defined at each
point in space in at least one of the gauges.  In this way the vector
potential can be considered a well defined connection on a
$U(1)$-bundle rather than a singular function\cite{WuYang}.  Although
this formal mathematical construction eliminates the singular Dirac
string, it also complicates calculations.  We will thus work in a
singular gauge being careful to remember --- when necessary --- that we
are really working with a more complicated mathematical object.

Another well known property of the monopole system is the Dirac
quantization condition\cite{Dirac} that tells us that the strength of
the magnetic monopole is quantized in units of the magnetic flux
quantum $\phi_0 = 2 \pi/e$ where $-e$ is the charge on an electron;
and, here and elsewhere in this paper, we have set $\hbar = c = 1$.

We will work in a gauge where the vector potential has no radial
component.  In spherical polar coordinates, one simple gauge choice is
the spherical analog of Landau gauge where we choose the vector
potential to point in the polar direction (ie, in the $\hat \phi$
direction, directed around the $\hat z$ axis):
\begin{equation}
  \vec A = \hat \phi A_{\phi}
\end{equation}
where
\begin{equation}
  A_{\phi}(\theta,\phi) = \frac{S}{e \sin \theta} (1-\cos \theta) .
\end{equation}
This gauge places the singular Dirac string at the south pole of the
sphere ($\theta = \pi$).  It is easy to check that the magnetic field
is given by $\vec B = \nabla \times \vec A = \frac{S}{e} \vO$ at all
points $\vO$ on the unit sphere except at the south pole where the
Dirac string cuts through the sphere.  The Dirac quantization
condition then dictates that $2S$ is an integer.

More generally, by rotating the above solution, we can construct a
gauge where the Dirac string is at an arbitrary point $\vO '$ on the
sphere.  The vector potential $\vec A_{\vO'}$ for such a gauge is
given by
\begin{equation}
  \label{eq:monovec}
  \vec A_{\vO'}(\vO) = \frac{2S}{e} \frac{ \vO \times \vO'} {|\vO -
    \vO'|^2} .
\end{equation}

\subsection{Defining The System}
\label{sub:Define}

We consider a monopole of magnetic charge $2S$ flux quanta at the
center of a unit sphere with $N$ electrons of charge $-e$ restricted
to the surface of the sphere.  The points on the unit sphere are
represented by unit vectors $\vO$ from the origin.  The length scale
of the system is determined by the magnetic length $l_0 =
(eB)^{-(1/2)}$ which can be altered by changing the charge on the
monopole. The Hamiltonian for this system can be written as
\begin{equation}
  \label{eq:H=T+V}
  H = T + V
\end{equation}
where the kinetic energy is given by
\begin{equation}
  T = \frac{1}{2m_b} \sum_{j=1}^N (\vec p_j +e \vec A(\vO_j))^2
\end{equation}
where $\vec p_j$ and $\vO_j$ are the momentum and position of the
$j^{th}$ particle, $\vec A$ is the vector potential, and $m_b$ is the
band mass of the electron.  The potential energy $V$ is given by
\begin{equation}
  \label{eq:firstquantV}
  V = \frac{1}{2} \sum_{i \ne j} v(\vO_i,\vO_j)
\end{equation}
where we will take the interaction to be a Coulombic $1/r$ potential
with $r$ the chord distance, ie
\begin{equation}
  v(\vO,\vO') = \frac{e^2}{\epsilon} \frac{1}{|\vO - \vO'|}
\end{equation}
where $\epsilon$ is the dielectric constant of the medium (which may be allowed
to change as a function of the magnetic length).  The restriction
of the system to a sphere is best represented by rewriting
\begin{equation}
  T = \frac{1}{2m_b} \sum_{j=1}^N |\vcalp_j + e \vcala(\vO_j)|^2
\end{equation}
where
\begin{eqnarray} \label{eq:vcalpdef}
  \vcalp_j &=& \vO \times \vec p_j \\ \vcala &=& \vO \times \vec A
  \label{eq:vcaladef}
\end{eqnarray}
In terms of this ``angular'' vector potential, the radial component of
the magnetic field $\vec B$ is given by
\begin{equation}
\label{eq:BdotR}
\vO \cdot \vec B = \vO \cdot (\vec \nabla \times \vec A) = -\vec \nabla
\cdot \vcala.
\end{equation}
The tangential component of the magnetic field does not couple to our
system of electrons confined to the sphere and can therefore be
ignored.

It is convenient to rewrite the kinetic and potential energies in
terms of the operator $\psi_e^+(\vO)$ that creates an electron at the
point $\vO$.  In this second quantized notation, we have
\begin{eqnarray}
  T&=& \frac{1}{2m_b} \,\,\int d\vO \, \psi_e^+(\vO)\,\, |\vhcalp + e
  \vcala(\vO) |^2 \,\,\psi_e(\vO) \\
  \label{eq:potential}
  V&=& \frac{1}{2} \int d\vO \int d\vO' \,\, v(\vO,\vO') \,\,
  :\den(\vO) \den(\vO'):
\end{eqnarray}
where the colons represent normal ordering of the operators, and
$\vhcalp$ is the canonical angular momentum operator
\begin{equation}
  \vhcalp = \vO \times \hatn p = \vO \times -i \vec \nabla,
\end{equation}
and $\den$ is the density given by
\begin{equation}
  \den(\vO) = \psi_e^+(\vO)\psi_e(\vO).
\end{equation}

\subsection{Chern-Simons Transformation}
\label{sub:CST}

As described originally by Lopez and Fradkin\cite{Lopez}, a singular
gauge transformation can be used to convert the electrons into
composite Fermions -- electrons bound to an even number $\tilde \phi =
2m$ of Chern-Simons flux quanta, where $m$ is an integer.  Such a
transformation provides a natural way of understanding many of the
features of the fractional quantized Hall
effect\cite{Simon,Tome,Lopez,Jain}.  To make such a transformation, we
define a vector valued Green's function on the sphere (see Eq.
(\ref{eq:monovec})) for two unit vectors $\vO$ and $\vO'$ which
corresponding to points on the sphere
\begin{equation}
\label{eq:tildeg}
\vec g(\vO,\vO') = - \frac{ \vO \times \vO'} {|\vO - \vO'|^2} +
\frac{\vO \times \vec S}{|\vO - \vec S|^2}
\end{equation}
where $\vec S$ is the unit vector pointing toward the south pole, and
the point $\vO = \vO'$ should be excluded from the Green's function.
The function $\vec g$ is the vector potential associated with an
infinitely thin magnetic flux tube carrying a single magnetic flux
entering the sphere at the south pole, and leaving the sphere at the
point $\vO'$.  Note that just as in the case of the magnetic monopole,
we can not write a vector potential that represents a flux quanta
entering the sphere but not leaving the sphere somewhere else.  As in
the case of the monopole, we have chosen the Dirac string to leave the
sphere through the south pole.  As we will see below, the Dirac string
through the south pole is an artifact of the gauge choice and can
essentially be ignored, whereas the flux quanta elsewhere through the
sphere will have important physical effects.

Defining an auxiliary (singular and multivalued) function
$f(\vO,\vO')$ such that
\begin{equation}
 \vec g(\vO,\vO') = \vec \nabla_{\vO} f(\vO,\vO'),
\end{equation}
we use the function $f$ (which we will never actually need to
evaluate) to define the quasiparticle creation operator
\begin{equation}
  \label{eq:psidef}
  \psi^+(\vO) = \psi^+_e(\vO) \exp \left[ -i \tilde \phi \int d\vO'
  f(\vO,\vO') \den(\vO') \right]
\end{equation}
which creates an electron bound to $\tilde \phi$ inward directed
Chern-Simons flux quanta at the point $\vO$, as well as creating a
Dirac string carrying the same flux out the south pole.  Although the
function $f$ is multivalued, the exponential in Eq. (\ref{eq:psidef})
is well defined and single valued since $\den(\vO')$ is a sum of
$\delta$-functions, and $e^{-i \tilde \phi f}$ is single valued.  As
in the case of the simple magnetic monopole, the string carrying flux
quanta out of the south pole is an artifact of our gauge choice and
has no physical effects as long as the net flux through the string
satisfies the Dirac quantization condition.  One should note however,
that although the flux entering the sphere at the position of the
quasiparticle looks very similar to the Dirac string, its position is
a dynamical variable.  We find that, unlike the Dirac string, these
flux quanta do in fact have physical effects.

 Since the density of electrons is equal to
the density of quasiparticles, the density operator is given by
\begin{equation}
  \den(\vO) = \psi_e^+(\vO)\psi_e(\vO) = \psi^+(\vO)\psi(\vO),
\end{equation}
and the equation for the potential energy (\ref{eq:potential})
remains unchanged.  On the other hand, in terms of the quasiparticle
operators, the kinetic energy is now written as
\begin{equation}
  \label{eq:HCSdef}
  T= \frac{1}{2m_b} \,\,\int d\vO \, \psi^+(\vO)\,\, |\vhcalp + e
  \vcala(\vO) - \vcalsa(\vO)|^2 \,\,\psi(\vO)
\end{equation}
where
\begin{eqnarray}  \label{eq:vcalsadef}
  \vcalsa(\vO) &=& \vO \times \vec a(\vO) \\ \vec a(\vO) &=& \tilde \phi
  \int d\vO'  \vec
  g(\vO,\vO') \den(\vO')
\end{eqnarray}
and $\vec a$ is the Chern-Simons vector potential.

\subsection{Mean-Field Theory}
\label{sub:MFT}

We begin our analysis by considering a simple mean field approach.
The total magnetic flux through the sphere seen by a
quasiparticle is an integer number $2S$ of quanta from the monopole
charge minus $\tilde \phi(N-1)=2m(N-1)$ Chern-Simons flux quanta bound
to the {\it other} quasiparticles.  (The quasiparticle does not see
the flux bound to itself).  As discussed above, the flux through the
Dirac string is static, and therefore should have no physical effects.
Thus, the flux through the string is not ``seen'' by the
quasiparticle.  We conclude that a given quasiparticle sees a total
field corresponding to a monopole charge of
\begin{equation}
\label{eq:DeltaSdef}
2 \Delta S = 2 S - 2m (N-1),
\end{equation}
flux quanta.  Within mean field theory, the associated magnetic field
is considered to be constant and uniform.  The effective Hamiltonian
for this mean field system is then given by
\begin{equation}
  \label{eq:H0def}
  H_0 = \frac{1}{2m_b} \,\,\int d\vO \, \psi^+(\vO)\,\, |\vhcalp + e
  \Delta \vcala(\vO)|^2 \,\,\psi(\vO)
\end{equation}
where $\Delta \vcala$ is the angular vector potential associated with
the magnetic monopole charge of $2 \Delta S$ quanta (see Eqs.
(\ref{eq:monovec}), (\ref{eq:vcaladef}), and (\ref{eq:DeltaSdef})),
and we have clearly neglected the Coulomb interaction.  This is simply
the Hamiltonian for a system of noninteracting quasiparticles on the
unit sphere in the field of a monopole of charge $2 \Delta S$ flux
quanta.  Such a problem has been solved previously\cite{WuYang,Tamm}
and the eigenstates for such a system are given by the monopole
spherical harmonics $Y^{(\Delta S)}_{lm}$ of Wu and Yang\cite{WuYang},
where where $l$ takes on the values $\Delta S, \,\, \Delta S+1, \,\,
\Delta S+2, \,\, \ldots \,$, and $m$ takes on the values $ -l,\,
-l+1,\, \ldots \, , l-1, \, l$.  The energies of these eigenstates are
given by
\begin{equation}
  \label{eq:omsubp}
  \omega_l \equiv E(l) = \frac{1}{2m_b} [l (l+1) - (\Delta S)^2]
\end{equation}
and thus the degeneracies are given by
\begin{equation}
  \mbox{Degeneracy}(l) = 2l+1.
\end{equation}
These $l$ states should be thought of as analogous to the usual Landau
levels on a plane.  We expect that when the number of electrons (and
thus the number of quasiparticles) is such that an integer number of
these quasiparticle Landau levels are filled, we will have an integer
quantized Hall state for the quasiparticles, and hence a stable
fractional quantized Hall state for the original electron
system\cite{Lopez,Jain}.  This condition occurs for $N=2\Delta S+1,
\, 4\Delta S+4, \ldots$.  The general formula is given by
\begin{equation}
  N= 2p \Delta S + p^2
\end{equation}
where $p$ is a positive integer.  Substituting in the definition
(\ref{eq:DeltaSdef}) of $\Delta S$, we find full quasiparticle Landau levels
and hence stable quantized Hall states when
\begin{equation}
  \label{eq:nupred}
  N = 2p (S-m(N-1)) + p^2.
\end{equation}
For an infinitely large system (the $S \rightarrow \infty$ limit) this
equation predicts quantized Hall  states at the Jain filling
fractions\cite{Jain}
\begin{equation}
  \nu(m,p) = \frac{N}{2S} = \frac{p}{2mp+1}
\end{equation}
which, along with their quasiparticle-quasihole conjugates, include
all of the experimentally well established  fractional quantized Hall states
within the first Landau level.  On the finite spherical system
however, Eq. (\ref{eq:nupred}) predicts that the quantized Hall
states occur when
\begin{equation}
  2S(N,m,p) = [\nu(m,p)]^{-1} N - (2m+p).
\end{equation}
The deviation ($2m+p$) from the infinite system filling fraction is a
finite size effect which has been called the ``Shift'' of the
state\cite{WenZee}.  The shifts predicted by this mean field composite
fermion model are in accordance with those predicted by Haldane's
formula for the hierarchy of stable states on the
sphere\cite{Haldane,Morf,Gros} and also agree with Jain's composite
fermion results on the sphere\cite{Jain}.

\section{The Response Function}
\label{sec:Response}

\subsection{Spherical Basis}
\label{sub:Basis}

In order to exploit the rotational symmetry of this system, we will
want to work in a basis that transforms in a simple way under
rotation.  Specifically, we will expand functions in angular momentum
eigenstates.  For scalar functions this can be done using the
usual\cite{Edmonds,Normand} spherical harmonics $Y_{lm}$.  However,
for vector valued functions, we will have to use a more complicated
vector function basis.  Furthermore, since our system is restricted to
the surface of a sphere, we will want to work with a basis that
naturally decouples the radial degree of freedom.  We define a
vector function basis\cite{Normand} as
\begin{eqnarray}
  \label{eq:T1def2}
  \TT{1lm} &=& \frac{ r \vec \nabla Y_{lm}}{\sqrt{l(l+1)}},
  \,\,\,\,\,\,\,\,\, l \ge 1,  \\
  \label{eq:T2def2}
  \TT{2lm} &=& \frac{-\vec r \times \vec \nabla Y_{lm}}{\sqrt{l(l+1)}},
  \,\,\, l \ge 1,
  \\ \TT{3lm} &=& \hatn r Y_{lm},  \,\,\,\,\,\,\,\,\,\,
  \,\,\,\,\,\,\,\,\,\,\, l\ge 0,
\end{eqnarray}
where $\vec r$ is the radial vector of magnitude $r$ and direction
(unit vector) $\hatn r$.  Note that the vector $\TT{3}$ is radial,
whereas $\TT{1}$ and $\TT{2} $ are tangential to the sphere. Thus, we
will be  able to completely decouple the $\TT{3}$ direction from our
problem.

This $\TT{}$ basis is orthonormal and complete on the space of
vector valued functions on the surface of a sphere.  That is to say
\begin{equation}
  \int d \vO \,\,  [\TT{ilm}(\vO)]^*\cdot[\TT{jkn}(\vO)] = \delta_{ij}
  \delta_{lk} \delta_{mn} \,\, ,
\end{equation}
and given a vector valued function $\vec f(\vO)$, we have
\begin{equation}
   \vec f(\vO) = \sum_{ilm} f_{ilm} \TT{ilm}(\vO),
\end{equation}
where
\begin{equation}
  f_{ilm} = \int d\vO \, [\TT{ilm}(\vO)]^* \cdot \vec f(\vO).
\end{equation}

Sometimes we will find it more convenient to work with
``four-vectors.''  To this end, we define a unit vector $\hatn e_w$
which is orthogonal to all three elements of the usual
three-dimensional Cartesian basis $\{\hatn e_x, \hatn e_y, \hatn e_z
\}$, where the metric is always taken to be Euclidean.  Any
four-vector can then be expressed in terms of the Cartesian basis and
the unit vector $\hatn e_w$.  In order to represent four-vector valued
functions on the sphere, we define a $0^{th}$ component of the $\TT{}$
basis as
\begin{equation} \label{eq:T4def}
  \TT{0lm}(\vO) = Y_{lm}(\vO) \hatn e_w
\end{equation}
which, combined with the other three $\TT{}$ basis vectors form an orthonormal
and complete basis set for functions on the surface of the sphere with
values in four dimensional space.  We will use the Greek indices $\mu$
and $\nu$ exclusively for such four-vectors, while the Latin indices
$i,j,..$ will be used for the indices of a three-vector.

\subsection{Definition of the Response Function}
\label{sub:DefofK}

We are concerned with finding the response of our system to electromagnetic
perturbations. The angular current operator is defined by
\begin{equation}
  \vec \Gamma^{\vcala} (\vO) = \frac{1}{e} \frac{\delta H}{\delta
    \vcala(\vO)} = \frac{1}{m_b} \psi_e^+(\vO) \left[\vhcalp + e\vcala
\right] \psi_e(\vO)
\end{equation}
Sometimes we will leave out the superscript $\vcala$ when it is clear
what the vector potential is in the system.  Note that there is no
radial component of the angular current.  We would like to think of
the fluctuation in the density as the $0^{th}$ element of this vector,
thus forming a four-vector. We define (with  $\vcala$ a dummy
index here)
\begin{equation}
  \label{eq:Gamma0def}
  \Gamma_0^{\vcala}(\vO) = \den(\vO) - \bar \den =
  \psi_e^+(\vO)\psi_e(\vO) - \bar \den
\end{equation}
where
\begin{equation}
  \bar \den = \frac{N}{4\pi.}
\end{equation}
Similarly, we would like to think of the scalar potential $-\Phi$ as
the $0^{th}$ element of the (angular) vector potential (see Eq.
(\ref{eq:vcaladef}))
\begin{equation}  \label{eq:cala0def}
   \cala_0 = -\Phi.
\end{equation}

If we now subject the system to an external time dependent electromagnetic
perturbation $\cala_\mu^{\mbox{\small ext}}$ the resulting linear response in
the angular current and density can be given in terms of a response
function $K_{\alpha \beta}(\vO,t;\vO',t')$ by the equation
\begin{equation} \label{eq:responseold}
  \left\langle \Gamma^{\vcala}_{\alpha}(\vO,t) \right\rangle = e
  \int_{-\infty}^{\infty} dt' \int d\vO' \,\,\, K_{\alpha
    \beta}(\vO,t;\vO',t') \cala_{\beta}^{\mbox{\small ext}}(\vO',t') .
\end{equation}
where the indices $\alpha$ and $\beta$ can take on the value $0$
(corresponding to the density and the scalar potential) or $1,2,3$
corresponding to the vector components of the current or vector
potential for some orthonormal vector basis.  Note that here $\vcala$
is the sum of the background vector potential from the monopole and
the perturbation $\cala_\mu^{\mbox{\small ext}}$.

We now use our $\vec T$ vector function basis to expand the quantities
in this equation:
\begin{eqnarray}
  \Gamma^{\vcala}_{\mu lm}(\omega) &=& \int dt \, e^{i\omega t} \int
  d\vO \, [\TT{\mu lm}(\vO)]^* \cdot \vec \Gamma(\vO) \\
  \label{eq:calaspheredef}
  \cala_{\mu lm}^{\mbox{\small ext}} (\omega) &=& \int dt \, e^{i\omega t} \int
d\vO \,
  [\TT{\mu lm}(\vO)]^* \cdot \vcala^{\mbox{\small ext}} (\vO)
\end{eqnarray}
where $\TT{}, \vec \Gamma$, and $\vcala$ are considered four-vectors
in these equations.  Note that $\Gamma_3 = \cala_3 = 0$, since $\vec
\Gamma$ and $\vcala$ are tangential vectors and $\TT{3}$ is radial.
Similarly, we can define
\begin{equation}
  K_{\mu l m; \nu l' m'}(\omega) = \int d(t-t') e^{i\omega
    (t-t')} \int d\vO \int d\vO' [\TT{\mu lm}(\vO)]^* \cdot
  K_{(\alpha,\beta)} (\vO,\vO', t-t') \cdot [\TT{\nu l'm'} (\vO')]
\end{equation}
where the first dot product is with respect to the $\alpha$ index of
$K$ and the second dot product is with respect to the $\beta$ index of
$K$.  In this equation we have explicitly used the time translational
invariance of the response function (ie, that  $K$ is a function of
$t-t'$ only).  In terms of these transformed quantities, the response equation
(\ref{eq:responseold}) is written in the simple form
\begin{equation}
  \label{eq:sphereresponse}
 \left\langle  \Gamma^{\vcala}_{\mu lm}(\omega) \right\rangle = e \sum_{\nu,
l', m'} K_{\mu l m; \nu l'm'}(\omega) \cala_{\nu l'm'}^{\mbox{\small ext}}
  (\omega).
\end{equation}
The advantage of having
made this transformation is that the spherical symmetry of the system
is now reflected in the simplifying relationship (proven in Appendix
\ref{app:rot}),
\begin{equation}
    \label{eq:Kdelta}
    K_{\mu l m;  \nu l'm'}(\omega) = \delta_{ll'}
    \delta_{mm'} K_{\mu \nu}(l,\omega)
\end{equation}
for $m = -l, \, -l+1, \, \ldots, \, l-1, \, l$.  In Appendix
\ref{app:cons} we use gauge invariance and current conservation to
derive further restrictions on the form of the response function
$K_{\mu \nu}(l,\omega)$.


\subsection{Bare Mean-Field Response}
\label{sub:MFR}

As a starting point, we will need to calculate the ``bare'' electromagnetic
response function $K^0_{\mu \nu}$ for the mean field system defined by the
Hamiltonian
$H_0$ given in Eq. (\ref{eq:H0def}).  This is simply the problem of
finding the response function for a system of $N$ noninteracting
quasiparticles of charge $-e$ on a unit sphere around a monopole of
charge $2\Delta S$ which creates a vector potential $\Delta \vcala$
such that an integer number of quasiparticle Landau levels are
filled.  If we add an external perturbing (angular) vector potential
$\vcala^{\mbox{\small ext}}$, the perturbation Hamiltonian is given by
\begin{equation}
  H_0^{\mbox{\small ext}} =  \int d\vO \,  \left[ e \vec
  \Gamma^{\Delta \vcala}(\vO) \cdot \vcala^{\mbox{\small ext}}(\vO)
 + \frac{e^2}{2m_b}  \den(\vO)
    \vcala^{\mbox{\small ext}}(\vO)  \cdot \vcala^{\mbox{\small ext}} (\vO)
\right]
\end{equation}
where the dot product in the first term (the paramagnetic term) is a
four-vector dot product whereas the dot product in the second term
(the diamagnetic term) is a three-vector dot product.  One must now
include the contribution of the perturbation Hamiltonian in the full
physical current:
\begin{equation}
\label{eq:diac}
  \vec \Gamma^{\Delta \vcala +\vcala^{\mbox{\small ext}} }(\vO) =
  \frac{1}{e} \frac{\delta (H_0 + H_0^{\mbox{\small ext}})}{\delta
    \vcala(\vO)} = \vec \Gamma^{\Delta \vcala} +
  \frac{e}{m_b}\den(\vO)  \vcala^{\mbox{\small ext}}(\vO).
\end{equation}
Using standard linear response theory\cite{FetterWalecka}, we now have
\begin{equation}
  \left\langle  \Gamma^{\Delta \vcala}_{\alpha} (\vO,t) \right
  \rangle_{H_0 + H_0^{\mbox{\tiny ext}}} = \left\langle
  \Gamma^{\Delta \vcala}_{\alpha} (\vO,t) \right \rangle_{H_0} - i
  \int_{-\infty}^{t} dt' \int d\vO' \left\langle \left[
  \Gamma^{\Delta \vcala}_{\alpha} (\vO,t), H_0^{\mbox{\small ext}}
\right] \right\rangle_.
\end{equation}
In the unperturbed state, there is no current, so the first term on
the right hand side of this equation vanishes.  Furthermore, since we
are working to linear order, the commutator of the diamagnetic term of
the perturbation Hamiltonian is neglected.  Thus we rewrite this
equation as
\begin{eqnarray}
  \left \langle  \Gamma^{\Delta \vcala+\vcala^{\mbox{\tiny ext}}}_{\alpha}
  (\vO,t) \right \rangle  &=& \frac{e\bar \den }{m_b} [1 - \delta_{\alpha 0}]
  \vcala^{\mbox{\small ext}}_{\alpha}(\vO,t) \\ &-& i e
  \int_{-\infty}^{t} dt' \int d\vO' \left\langle \left[
  \Gamma^{\Delta \vcala}_{\alpha} (\vO,t), \vec \Gamma^{\Delta
    \vcala}(\vO',t') \right] \right\rangle \cdot \vcala^{\mbox{\small
      ext}}(\vO',t')
\end{eqnarray}
where we have used Eq. (\ref{eq:diac}) to obtain the first term on the
right hand side.  Thus the response function is written as the sum of
diamagnetic and paramagnetic parts as
\begin{equation}
  K^0_{\alpha \beta} (\vO,t; \vO',t') = \frac{e \bar \den}{m_b}
  \delta(\vO - \vO') \delta(t - t') [1-\delta_{\beta 0}]
  \delta_{\alpha \beta} + D^0_{\alpha \beta} (\vO,t; \vO',t')
\end{equation}
where the superscript $0$ indicates that it is a mean-field quantity,
any diamagnetic current in the radial direction is projected out, and
the paramagnetic piece is just the retarded current correlation
function given by
\begin{equation}
  \label{eq:D0def}
  D^0_{\alpha \beta}(\vO,t; \vO',t')  = - i \theta(t-t') \left\langle
  \left[ \Gamma^{\Delta \vcala}_{\alpha}(\vO,t), \Gamma^{\Delta
    \vcala}_{\beta}(\vO',t') \right] \right\rangle_.
\end{equation}
In terms of angular momentum and frequency components, we have
\begin{equation}
  K^0_{\mu lm; \nu l' m'}(\omega) = D^0_{\mu lm; \nu l'm'}(\omega) +
  \frac{e \bar \den}{m_b} \delta_{\mu \nu} \delta_{l l'} \delta_{m m'}
  [1 - \delta_{\mu 0}]
\end{equation}
where
\begin{equation}  \label{eq:Dsphere}
 D^0_{\mu lm; \nu l'm'}(\omega)  =  \int
d(t-t')  \int d\vO \int d\vO'   e^{i\omega (t-t')}
[\TT{\mu lm}(\vO)]^* \cdot    D^0_{(\alpha \beta)}(\vO, \vO',t-t')
 \cdot [\TT{\nu l'm'} (\vO')],
\end{equation}
where the first dot product is with respect to $\alpha$ and the second
dot product is with respect to $\beta$, and all the vectors are
four-vectors.  Once again, we expect that the spherical symmetry (see
Appendix \ref{app:rot}) will allow us to write the retarded correlation
function as
\begin{equation}
   D^0_{\mu lm; \nu l'm'}(\omega) =  \delta_{ll'} \delta_{mm'} D^0_{\mu
\nu}(l,\omega)
\end{equation}
so that the bare response function may be written as
\begin{equation}
  \label{eq:K0final}
   K^0_{\mu \nu}(l,\omega) = D^0_{\mu  \nu}(l,\omega) + \frac{
 e  \bar \den}{m_b} \delta_{\mu \nu} [1 - \delta_{\mu 0}]
\end{equation}
The actual calculation of the retarded correlation function $D^0$ is
relegated to Appendix \ref{app:ret}.


\section{The RPA}

\subsection{Self Consistent RPA}
\label{sub:SCRPA}

The RPA on a plane can be defined by saying that the quasiparticles
respond via the bare response function $K^0$ to an effective scalar
and vector potential $ A^{\mbox{\small eff}}_{\mu}$ given by
\begin{equation}
 A^{\mbox{\small eff}}_{\mu} = A^{\mbox{\small ext}}_{\mu} +
A^{\mbox{\small in}}_{\mu}
\end{equation}
where $A^{\mbox{\small in}}_{\mu}$ is an induced vector potential
which includes a contributions from a self consistently calculated
Coulomb potential due to the induced variation of the electron density
and a contribution arising from the self consistent Chern-Simons
magnetic and electric fields given by\cite{Lopez}
\begin{eqnarray}
  \vec b^{\mbox{\small
    in}}(\vec r,t) &=& 2 \pi \tilde \phi \left[ \left\langle  \den(\vO,t)
\right\rangle - \bar \den \right]  \hatn z \\
  \vec e^{\mbox{\small in}} (\vec r,t) &=& 2 \pi \tilde \phi \, \hatn z \times
\left\langle \vec J \right\rangle
\end{eqnarray}
where $\langle \den \rangle$ and $\langle \vec J \rangle$ are the
density and current calculated self consistently in the RPA.  For the
present case of $N$ electrons on a sphere, we write the analogous
expressions
\begin{eqnarray}
  \label{eq:CSb}
  \vec b^{\mbox{\small
    in}}(\vO,t) &=&
    {\scriptstyle \left( \frac{N-1}{N} \right)} \, 2 \pi \tilde \phi \,
\left\langle \Gamma_0(\vO,t) \right\rangle \vO =
{\scriptstyle \left( \frac{N-1}{N} \right)} \,2 \pi  \tilde \phi \, \left[
\left\langle  n_e(\vO,t) \right\rangle - \bar \den \right] \,  \vO \\
\label{eq:CSe}
  \vec e^{\mbox{\small
    in}}(\vO,t) &=& {\scriptstyle \left( \frac{N-1}{N} \right)} \, 2 \pi \tilde
\phi \, \left\langle \vec \Gamma(\vO,t) \right\rangle
\end{eqnarray}
where $\left\langle \Gamma_0 \right \rangle$ and $\left\langle \vec
\Gamma \right \rangle$ are the induced density fluctuation and angular
current calculated self consistently in the RPA.  The factor of
$\frac{N-1}{N}$ is introduced here to correct for the fact that the
quasiparticle does not see its own field.  This factor disappears in
the limit of $N \rightarrow \infty$, but is clearly necessary if we
wish to recover the correct response function for a single electron
when $N=1$.  Similarly, the induced Coulomb potential is taken to be that
arising from a charge density
\begin{equation}
  \delta \den^{\mbox{\small
    in}}(\vO,t) =  {\scriptstyle \left( \frac{N-1}{N} \right)} \,
\left\langle \Gamma_0(\vO,t) \right\rangle = {\scriptstyle \left(
  \frac{N-1}{N} \right)}  \, \left[ \left\langle
n_e(\vO,t) \right\rangle - \bar \den \right]
\end{equation}
so that an induced electric field $\vec E^{\mbox{\small
    in}}$ can be defined whose three-dimensional divergence satisfies
\begin{equation}
  \label{eq:Coulin}
  4 \pi \epsilon \vec \nabla \cdot  \vec E^{\mbox{\small in}}(\vec r) = \delta
  \den^{\mbox{\small in}}(\vec r/|\vec r|) \,  \delta(|\vec r| -1).
\end{equation}

The required induced Chern-Simons fields and Coulomb potential can be
obtained on the sphere from an induced angular vector potential
$\vcala^{\mbox{\small in}} = \vO \times \vec A^{\mbox{\small in}}$ and
scalar potential $\cala_0^{\mbox{\small in}} = A^{\mbox{\small in}}_0$
whose angular momentum components are given by
\begin{equation}
  \label{eq:Ain}
  e \cala_{\mu lm}^{\mbox{\small in}} = {\scriptstyle \left(
  \frac{N-1}{N} \right)} \sum_{\nu} U_{\mu \nu}(l) \left \langle
  \Gamma_{\nu lm} \right \rangle
\end{equation}
where
\begin{equation}
  \label{eq:Unew}
  U_{\mu \nu}(l) = v(l) \delta_{\mu 0} \delta_{\nu 0} + w(l) \left[
  \delta_{\mu 1} \delta_{\nu 0} + \delta_{\nu 1} \delta_{\mu 0}
\right]
\end{equation}
and the coefficients $v$ and $w$ are given by
\begin{eqnarray}
  v(l) &=&
 \frac{e^2}{\epsilon} \frac{4\pi}{2l+1} \\
  w(l) &=& \tilde \phi  \frac{2\pi}{\sqrt{l(l+1)}}   \,\,.
\end{eqnarray}

To derive Eqs.  (\ref{eq:Ain}) and (\ref{eq:Unew}) we calculate the
scalar potential and angular vector potential created by the $N$
quasiparticles (charge $-e$ particles attatched to $\tilde \phi$ flux
quanta) when their density and current is given by $\left\langle
\Gamma_{\mu} \right\rangle$.  These potentials are then multiplied by
the above mentioned factor of $\frac{N-1}{N}$ to account for the fact
that a quasiparticle does not see its own field. With this
presciption, the Coulomb contribution to the scalar potential from
induced charge fluctuation is given by
\begin{equation}
  \label{eq:intCoul}
 e \cala^{\mbox{\small in}}_0(\vO)_{\mbox{\small Coulomb}} =
  {\scriptstyle \left(
  \frac{N-1}{N} \right)} \, \int d\vO' \, v(\vO,\vO')  \langle \Gamma_0(\vO')
\rangle
\end{equation}
whereas the scalar potential from the motion of the Chern-Simons flux quanta is
given by
\begin{equation}
\cala^{\mbox{\small in}}_0(\vO)_{\mbox{\small Chern-Simons}} =
  {\scriptstyle \left(
  \frac{N-1}{N} \right)} \, \int d\vO' \, \vec W(\vO,\vO')  \cdot \langle \vec
\Gamma(\vO') \rangle
\end{equation}
so that
\begin{equation}
  \cala^{\mbox{\small in}}_0(\vO) = \cala^{\mbox{\small
in}}_0(\vO)_{\mbox{\small Coulomb}} + \cala^{\mbox{\small
in}}_0(\vO)_{\mbox{\small Chern-Simons}}
\end{equation}
and  the angular vector potential from the Chern-Simons
flux quanta is given by
\begin{equation}
  \vcala^{\mbox{\small in}}(\vO) =  {\scriptstyle \left(
  \frac{N-1}{N} \right)}\,  \tilde \phi \,
  \int d\vO' \, \vec  W(\vO,\vO')  \langle \Gamma_0(\vO') \rangle
\end{equation}
where
\begin{equation}
\vec  W(\vO,\vO') =  - \frac{\vO \times (\vO
    \times \vO')} {|\vO - \vO'|^2.}
\end{equation}
is the flux quanta Green's function (See Eq. (\ref{eq:tildeg})).
Finally, we use Eqs. (\ref{eq:vl}) and (\ref{eq:Wres}) to evaluate the
angular momentum components of $v$ and $\vec W$, and thus establish
the given form ((Eqs.  (\ref{eq:Ain}) and (\ref{eq:Unew})) of the
induced vector potential.

We can verify that this prescription (described by Eq. (\ref{eq:Ain}))
does in fact correspond to the induced Chern-Simons electric and
magnetic fields given by Eqs.  (\ref{eq:CSb}) and (\ref{eq:CSe}) as
well as the induced Coulombic electric field described by Eq.
(\ref{eq:Coulin}) by reconstructing the induced vector potential
\begin{eqnarray} \label{eq:Arec0}
  A_0^{\mbox{\small in}} &=&\cala^{\mbox{\small in}}_0 = \sum_{lm} \cala_{0
lm}^{\mbox{\small in}} \TT{0lm} \\ \label{eq:Arec}
  \vec A^{\mbox{\small in}} &=& -\vO \times \vcala^{\mbox{\small in}} =
\sum_{ilm} \cala_{i lm}^{\mbox{\small in}} \left(-\vO \times \TT{ilm} \right)
\end{eqnarray}
and then differentiating to find
\begin{eqnarray}  \label{eq:AeE}
\vec \nabla A^{\mbox{\small
      in}}_{0} - \frac{\partial}{\partial t} \vec A^{\mbox{\small in}}
  &=& \vec E^{\mbox{\small in}} + \vec e^{\mbox{\small
      in}}    \\ \label{eq:Ab}
  \vO \cdot (\vec \nabla \times \vec A^{\mbox{\small in}})       &=&
 \vec b^{\mbox{\small
      in}}
\end{eqnarray}
These equalities are demonstrated explicitly in Appendix \ref{app:EB}.

Once we have derived Eqs. (\ref{eq:Ain}) and (\ref{eq:Unew}) we can
make the RPA approximation that the quasiparticles respond via the
bare response to the effective angular vector and scalar potential
\begin{equation}
  \left\langle \Gamma_{\mu lm} \right\rangle = e \sum_{\nu} K^0_{\mu
    \nu}(l) \cala_{\nu lm}^{\mbox{\small eff}},
\end{equation}
where the effective angular vector and scalar potential is given by
the sum of  external and induced contributions
\begin{equation}
  \label{eq:Anew}
  \cala_{\mu}^{\mbox{\small eff}} = \cala_{\mu}^{\mbox{\small ext}} +
\cala_{\mu}^{\mbox{\small in}}
\end{equation}
On the other hand, the full response is defined via
\begin{equation}
  \left\langle \Gamma_{\mu lm} \right\rangle = e \sum_{\nu} K_{\mu
    \nu}(l) \cala_{\nu lm}^{\mbox{\small ext}}.
\end{equation}
Using Eq. (\ref{eq:Ain}) and eliminating $\langle \vec \Gamma \rangle$
we solve to find the RPA result
\begin{equation}
  \label{eq:RPA}
  K(l) =  K^0(l) \left[ 1 - {\scriptstyle  \left(\frac{N-1}{N} \right)}  U(l)
K^0(l) \right]^{-1}
\end{equation}
where all of these quantities are $3 {\scriptstyle{\times}} 3$
matrices, ($\mu, \nu = 0,1,2$).

Although this self consistent calculation has been performed in terms
of the response function $K$ and the full physical current $\vec
\Gamma$ it is also possible\cite{ChenFetter} to perform a self
consistent calculation in terms of the retarded correlation function
$D$ and the mean field current $\vec \Gamma^{\Delta \vcala}$.  The
result of such an approximation must then be corrected to account for
the fluctuations in the diamagnetic current.  However, once this
correction has been properly taken into account to find the response
$K$, the end result will be exactly the same as the RPA equation we
have given here.

At this point we note that there are restrictions on the response
function from gauge invariance (Eq. (\ref{eq:gaugeinvariance})) and
current conservation (Eq.  (\ref{eq:currentconservation})) derived in
Appendix \ref{app:cons}.  With these restrictions, along with the
manifest symmetries of the response matrix, there are only three
independent entries in the $3 {\scriptstyle{\times}} 3$ response
matrix, and we can write
\begin{equation}
 K(l,\omega)  = \left( \begin{array}{ccc}
 K_{00}(l,\omega) &   K_{01}(l,\omega) & \frac{-i\omega}{\sqrt{l(l+1)}}
K_{00}(l,\omega)  \\
 K_{01}(l,\omega) &   K_{11}(l,\omega) &\frac{-i\omega}{\sqrt{l(l+1)}}
K_{01}(l,\omega) \\
 \frac{i\omega}{\sqrt{l(l+1)}} K_{00}(l,\omega)  &
\frac{i\omega}{\sqrt{l(l+1)}} K_{01}(l,\omega)  & \frac{\omega^2}{l(l+1)}
K_{00}(l,\omega)
\end{array}
\right)_.
\end{equation}
Furthermore, the Hermitian interaction matrix $U$, considered as a $3
{\scriptstyle{\times}} 3$ matrix only has entries in the $(0,0),(0,1)$
and $(1,0)$ places.  If we choose the gauge of our external magnetic
field so that $\cala_{2lm}^{\mbox{\small ext}} =0$, we can eliminate
redundancies and reduce the RPA to a $2 {\scriptstyle{\times}} 2$
matrix equation.  In this gauge, the RPA equation (\ref{eq:RPA})
remains the same, but now the matrices are now $2
{\scriptstyle{\times}} 2$ and are given by
\begin{equation}
 K(l,\omega)  = \left( \begin{array}{cc}
 K_{00}(l,\omega) &   K_{01}(l,\omega) \\
 K_{01}(l,\omega) &   K_{11}(l,\omega)
\end{array}
\right)
\end{equation}
and
\begin{equation}
U(l) =  V(l) + W(l)
\end{equation}
where
\begin{eqnarray}
 V(l) &=&  \left( \begin{array}{cc}
 v(l)  &   0 \\
 0  &   0
\end{array}
\right)   \\
 W(l) &=& \left( \begin{array}{cc}
 0 & w(l) \\
w(l) & 0
\end{array}
\right)_.
\end{eqnarray}

It is sometimes convenient to think in terms of the conductivity
rather than the electromagnetic response\cite{Tome,Simon}.  The
conductivity $\sigma$ is defined as the response to the total
electromagnetic field $\cala_{\mu lm}$ whereas the electromagnetic
response $K$ is the response to the external electromagnetic field
$\cala_{\mu lm}^{\mbox{\small ext}}$.  The magnetic field generated by
the quantum Hall system is small, so there is essentially no
difference between $\vcala_{lm}$ and $\vcala_{lm}^{\mbox{\small
    ext}}$.  On the other hand, the scalar potentials $\cala_{0lm}$
and $\cala_{0lm}^{\mbox{\small ext}}$ differ by the Coulomb potential
$ v(l) \Gamma_{0lm}$ generated by density fluctuations.  Thus, we
define a $2 {\scriptstyle{\times}} 2$ matrix $\Pi(l,\omega)$ which is
more closely related to the conductivity, to be the electromagnetic
response without the Coulomb contribution:
\begin{equation}
  K^{-1}(l,\omega) = \Pi^{-1}(l,\omega) - {\scriptstyle \left(
    \frac{N-1}{N} \right) } V(l,\omega)_.
\end{equation}
In other words, $\Pi$ is the sum of all diagrams that are irreducible with
respect to the Coulomb interaction.

We now can define the $2 {\scriptstyle{\times}} 2$ conductivity matrix
$\sigma$ as
\begin{equation}
  \label{eq:convsigma}
  \sigma = i T \Pi T^*.
\end{equation}
where $T$ is the conversion matrix
\begin{equation}
  T = e \left( \begin{array}{cc}
   i \sqrt{\frac{\omega}{l(l+1)}}    &  0  \\
   0     &    \sqrt{\frac{1}{\omega}}    \end{array} \right)_.
\end{equation}
This conductivity matrix has been  constructed to satisfy Ohm's law
\begin{equation}
  -e \left( \begin{array}{c} \Gamma_{2lm} \\ \Gamma_{1lm} \end{array} \right) =
  \sigma
  \left( \begin{array}{c} \cale_{2lm} \\ \cale_{1lm} \end{array} \right)
\end{equation}
where $\vcale_{ilm}$ are the angular momentum components in the $\TT{}$ basis
of the angular electric field $\vcale$ which is given in terms of the
true electric field $\vec E$ as
\begin{equation}
  \label{eq:vcaledef}
   \vcale = \vO \times \vec E.
\end{equation}
In terms of angular momentum and frequency components (in our chosen
gauge) we can easily derive
\begin{eqnarray}
   \cale_{1lm} &=&  i \omega \cala_{1lm} \\
   \cale_{2lm} &=& - \cala_{0lm}\sqrt{l(l+1)}
\end{eqnarray}
which is then used, along with the gauge invariance
(\ref{eq:gaugeinvariance}) and current conservation (\ref{eq:precur}),
to derive the form of the conversion equation (\ref{eq:convsigma}).

Similarly, we can  define a bare quasiparticle conductivity $\tilde
\sigma$ which in the RPA we approximate as
\begin{equation}
  \label{eq:sigmaconvtilde}
  \tilde \sigma = i T K^0 T^*.
\end{equation}
It should be noted that these definitions of the conductivity in some
sense do not give the proper $\omega \rightarrow 0$
limit\cite{Tome,Simon}.  However, we will not be concerned with this
limit in the present paper, and we expect that over the range of
frequencies where we expect the RPA to be a good approximation, these
definitions are appropriate.

In terms of these conductivities the RPA equation is written as
\begin{eqnarray}
  \label{eq:Kfinal}
   K^{-1} &=& i T^* \rho T  - {\scriptstyle \left(
    \frac{N-1}{N} \right) } V  \\
   \label{eq:rhorhorho}
   \rho &=& \tilde \rho + \rho_{cs}  \\
   \rho_{cs} &=& i [T^{-1}]^*  W [T^{-1}]  =  \frac{2 \pi \tilde \phi}{e^2}
{\scriptstyle \left( \frac{N-1}{N}\right)} \left[ \begin{array}{cc}
 0 & -1 \\ 1 & 0 \end{array} \right]_.
\end{eqnarray}
where
\begin{eqnarray}
  \rho &=& \sigma^{-1} \\
\label{eq:sigmarhotilde}
  \tilde \rho &=& \tilde \sigma^{-1}.
\end{eqnarray}

\subsection{Mass Renormalization: Modified RPA}
\label{sub:Mass}

As pointed out in previous work\cite{Tome,Simon}, within the RPA
theory considered thus far, the quasiparticle effective mass $m^*$ is
just the bare band mass $m_b$.  In this theory, we perturb around the
reference Hamiltonian (\ref{eq:H0def}) that describes quasiparticles
with the bare band mass $m_b$. We expect however, that interactions
may renormalize the quasiparticle mass significantly.  In the limit
where the Coulomb interaction is turned off, all of the states in a
Landau level become degenerate, and the effective mass diverges.  In
order to estimate the importance of the mass renormalization when the
Coulomb interaction is turned back on, we follow Halperin, Lee, and
Read\cite{Tome} and make a crude estimate for the value of the
effective mass.  If we assume that the characteristic Coulomb
interaction energy ($e^2 / \epsilon l_0$) is small compared to the
characteristic energy spacing between Landau levels ($\hbar
\omega_c$), then level mixing can be neglected and all energies of
interaction should scale as $e^2 / \epsilon l_0$.  Thus, if a finite
effective mass $m^*$ exists, we should have
\begin{equation}
  \frac{\hbar^2}{m^* l_0^2}  \propto  \frac{e^2}{\epsilon l_0}
\end{equation}
or
\begin{equation}
\label{eq:mstar1}
  \frac{\epsilon}{e^2 l_0} = C m^*
\end{equation}
for some proportionality constant $C$ where we have set $\hbar = 1$
again.  The estimate, $C \approx 0.3$ has been made by Halperin, Lee,
and Read\cite{Tome} by examining results from exact diagonalization of
small systems.  Using the experimentally relevant dielectric constant
$\epsilon = 12.6$ appropriate for GaAs, a magnetic field $B=10T$, and
a filling fraction $\nu =\frac{1}{2}$, this estimate yields
\begin{equation}
  m^* \approx 4m_b.
\end{equation}

Using a self-consistent analysis of a selected set of diagrams for the
self energy of the transformed fermions which describes the
interaction with long wavelength fluctuations in the Chern-Simons
vector potential, Halperin, Lee, and Read \cite{Tome} conclude that
for the case of the Coulomb interaction between the electrons, the
effective mass $m^*$ should actually exhibit a logarithmic divergence
for energies near the Fermi energy, and for $p \rightarrow \infty$
(ie, for $\nu \rightarrow \frac{1}{2}$).  The coefficient in front of
the logarithm obtained by Ref. \cite{Tome} is relatively small,
however, and the resulting values of the effective mass, in practice,
will not be very different from those given by Eq.
(\ref{eq:mstar1}).

The important thing to note here is that the effective mass can be
renormalized considerably.  Therefore, our RPA procedure of expanding
around a Hamiltonian that describes particles of mass $m_b$ is likely
to be a poor approximation. As noted in Ref.  \cite{Simon}, if one
simply inserts the effective mass $m^*$ in place of the band mass
$m_b$ in the reference Hamiltonian (\ref{eq:H0def}), one obtains a
``na\"{\i}ve response'' function that should reasonably represent the
low energy excitations of the system\cite{Simon} but which violates
the $f$-sum rule.  In this section we will follow Ref. \cite{Simon}
and propose a ``Modified RPA,'' generalized to the sphere, that
reasonably represents the low energy excitations while satisfying the
$f$-sum rule\cite{end1}.

As described in Ref. \cite{Simon}, the
$f$-sum rule requires that the high frequency response of our system
must be determined by the bare band mass $m_b$ and not the effective
mass $m^*$.  More specifically, in the high frequency limit, the
response of $N$ interacting electrons on a sphere around a monopole of
charge $2S$ flux quanta  must be the same as the response of a system of $N$
noninteracting electrons on the same sphere.  In terms of the
resistivity matrix $\rho$, this limit is given as
\begin{equation}
  \label{eq:fsum1}
 \rho \sim \frac{m_b}{e^2 \bar n_e}
   \left[ \begin{array}{cc} -i \omega & \frac{-S}{m_b}  \\
         \frac{S}{m_b} &  - i \omega \end{array} \right]
\end{equation}
which can be verified by numerically examining the result obtained  in
Appendix \ref{app:ret} where we derive the response for a system of
noninteracting electrons (In the appendix, the total field is called
$2 \Delta S$).  Since at high frequency, the particles essentially
oscillate in place, it is not surprising that this high frequency
limit yields exactly the same high frequency resistivity as the
analogous planar system\cite{Simon}.

Now, in our Chern-Simons approach, we actually begin by calculating
the response of a system of quaisiparticles in the effective field of
a monopole of charge $2 \Delta S$ flux quanta.
In the RPA, the high
frequency limit of the quasiparticle resistance is given by
\begin{equation}
  \label{eq:qpfsum}
 \tilde \rho \sim \frac{m_b}{e^2 \bar  n_e}
   \left[ \begin{array}{cc} -i \omega & \frac{-\Delta S}{m_b}  \\
         \frac{\Delta S}{m_b} &  -i \omega \end{array} \right]_.
\end{equation}
as required by the $f$-sum rule for the effective quasiparticle
system.  This is then converted into a total resistance using Eq.
(\ref{eq:rhorhorho}) to yield
\begin{equation}
  \rho  \sim
  \frac{m_b}{e^2 \bar  n_e}
   \left[ \begin{array}{cc} -i \omega & \frac{-\Delta S}{m_b}  \\
         \frac{\Delta S}{m_b} &  -i \omega \end{array} \right]
  + \frac{2 \pi \tilde \phi}{e^2}
{\scriptstyle \left( \frac{N-1}{N}\right)} \left[ \begin{array}{cc}
 0 & -1 \\ 1 & 0 \end{array} \right]
\end{equation}
which can be shown to be equivalent to Eq. (\ref{eq:fsum1}) by using
relation (\ref{eq:DeltaSdef}) to relate $S$ to $\Delta S$.  More
generally, we see that our original electron system will satisfy the
$f$-sum rule with respect to the monopole charge $S$, (\ref{eq:fsum1}),
if and only if the quasiparticle system satisfies the $f$-sum rule
with respect to the effective monopole charge $\Delta S$, (\ref{eq:qpfsum}).

 As we mentioned above, we
know that the quasiparticle mass is renormalized to some new effective
mass $m^*$.  In the  na\"{\i}ve approach we  simply
replace the band mass $m_b$ by the effective quasiparticle mass $m^*$
everywhere it occurs in the calculation of $\tilde \rho$, which
amounts to simply replacing the band mass by the effective mass in the
Hamiltonian (\ref{eq:H0def}).  The result of such a substitution is
what we will call the na\"{\i}ve quasiparticle resistivity tensor
\begin{equation}
\tilde \rho^{\mbox{\small n}}  \sim \frac{m^*}{e^2 \bar  n_e}
   \left[ \begin{array}{cc} -i \omega & \frac{-\Delta S}{m^*}  \\
         \frac{\Delta S}{m^*} &  -i \omega \end{array} \right]
\end{equation}
which clearly violates the $f$-sum rule.  In order to repair this
na\"{\i}ve approximation, we will simply adopt the result of the
Landau Fermi liquid theory approach developed in Ref. \cite{Simon}.
In that paper, an analogous theory is developed in detail for a planar
geometry.  It is found that Fermi liquid theory can be used to self
consistently account for an arbitrary quasiparticle mass
renormalization.  The full resistivity is then given in
terms of the na\"{\i}ve resistivity and the mass renormalization as
\begin{equation}
  \label{eq:FLT1}
  \tilde \rho = \tilde \rho^{\mbox{\small n}} + \frac{i\omega(m^* -
    m_b)}{\bar n_e e^2} 1.
\end{equation}
If we assume that the same prescription works on the sphere, we find
that the resulting resistivity satisfies the $f$-sum rule as desired.
One can in fact formally derive this result on a sphere, following
the exact same procedure as given in Ref. \cite{Simon}.  The validity
of such an approach for small systems is, however, questionable.
Instead, we simply take Eq. (\ref{eq:FLT1}) as a motivated ansatz for
modifying our na\"{\i}ve results such that they satisfy the $f$-sum
rule and give the correct energy scale for low energy excitations.
In addition, this approach clearly agrees with Ref. \cite{Simon} in
the limit where the sphere is taken to be large.

The complete prescription for the modified RPA for quantized Hall
states on a sphere is to calculate the unperturbed response function
for the quasiparticle system $K^0$ for quasiparticles of mass $m^*$ as
described in Eqs. (\ref{eq:K0final}) and (\ref{eq:xxthis}) where all
occurrences of the band mass $m_b$ are replaced with the effective mass
$m^*$,  then convert to the naive resistivity $\tilde
\rho^{\mbox{\small n}}$ using Eqs. (\ref{eq:sigmaconvtilde}) and
(\ref{eq:sigmarhotilde}), and then add the off diagonal Chern-Simons term and
the diagonal mass renormalization term to yield the total resistivity
\begin{equation}
  \rho = \tilde \rho^{\mbox{\small n}} - \frac{i\omega (m_b-m^*)}{n_e
    e^2} \left[ \begin{array}{cc} 1 & 0 \\ 0 & 1 \end{array} \right] +
  \frac{2\pi \tilde \phi} { e^2 } {\scriptstyle \left(
    \frac{N-1}{N}\right)} \left[ \begin{array}{cc} 0 & -1 \\ 1 & 0
\end{array} \right]
\end{equation}
which can be converted back to a response function using Eq. (\ref{eq:Kfinal}).

\section{Summary}
\label{sec:summary}

In this paper we have used the fermion Chern-Simons approach to model
a fractional quantized Hall state on a sphere as an integer quantized
Hall state of transformed fermions interacting with a Chern-Simons
field.  Linear response theory is used to find the bare mean field response
function.  We then use the RPA to account for fluctuations in the
Chern-Simons field as well as the direct Coulomb interaction.  The RPA
result, however, can not properly account for the effects of mass
renormalization.  Using results derived in Ref. \cite{Simon} we
propose a ``modified RPA'' prescription that more appropriately
accounts for these mass renormalization effects.  A quantitative
comparison between the response functions derived in this paper and
response functions given by exact numerical diagonalizations will be
made in a following paper\cite{Song}.



\vspace{10pt}

{\centerline{\small
    ACKNOWLEDGEMENTS}}

The authors gratefully acknowledge helpful discussions with Song He.
This work was supported by the National Science Foundation Grant
DMR-91-15491.
\appendix
\setcounter{section}{0}
\renewcommand{\theequation}{\Alph{section}.\arabic{equation}}
\setcounter{equation}{0}

\section{Rotational Invariance in the Spherical Basis}

\label{app:rot}
Given a function $K_{\alpha \beta}(\vO,\vO')$ (such as our response
function) that is rotationally invariant, for any rotation $\tilde R$
around the origin of our three dimensional space, we have
\begin{equation}
   K_{(\alpha,\beta)} (\vO,\vO', t-t') =  K_{(\tilde R \alpha,\tilde R \beta)}
(\tilde R \vO,\tilde R \vO', t-t')
\end{equation}
where the $\tilde R$ applied to the $\alpha$ and $\beta$ indices
rotates the three-vector valued part of the $K$ function.
We now transform into the $\TT{}$ basis
\begin{eqnarray}
  K_{\mu lm;\nu l'm'} &=&  \int d\vO \int d\vO' [\TT{\mu lm}(\vO)]^* \cdot
  K_{(\alpha,\beta)} (\vO,\vO', t-t') \cdot [\TT{\nu l'm'} (\vO')].
\\  &=& \int d\vO \int d\vO' [\TT{\mu lm}(\vO)]^*
  \cdot K_{(\tilde R \alpha,\tilde R \beta)} (\tilde R \vO,\tilde R
  \vO', t-t') \cdot [\TT{\nu l'm'} (\vO')]
\end{eqnarray}
or equivalently
\begin{equation}
 K_{\mu lm;\nu l'm'} =  \int d\vO \int d\vO' \left[ \tilde R^{-1} \TT{\mu
lm}(\tilde R^{-1} \vO)^* \right] \cdot
  K_{(\alpha,\beta)} (\vO,\vO', t-t') \cdot \left[ \tilde R^{-1} \TT{\nu l'm'}
(\tilde R^{-1} \vO') \right]
\end{equation}
where we have used the fact that the measure $d\vO$ is invariant under
rotation.  Now, since the basis vectors $\TT{ilm}$ are spherical
tensors\cite{Edmonds,Normand} of rank $l$, under an arbitrary rotation
$\tilde R$, we have the rotation law
\begin{equation}
   \tilde R \TT{ilm}(\tilde R \vO)  = \sum_n D^l_{nm}(\tilde R) \TT{iln}
\end{equation}
where  $D$ is the rotation matrix as defined in
Ref. \cite{Normand}.  This then results in the identity
\begin{equation}
  K_{\mu lm; \nu l'm'} =  \sum_{n\, n'} [D^l_{nm}(\tilde
R)]^*[D^{l'}_{n'm'}(\tilde R)]
 K_{\mu ln; \nu l'n'}
\end{equation}
which must be true for all rotations $\tilde R$.  Thus we can
integrate over all possible rotations and use the orthogonality
relation\cite{Edmonds,Normand}
\begin{equation}
   \int d\tilde   R \, [D^l_{nm}(\tilde R)]^*[D^{l'}_{n'm'}(\tilde R)] =
\frac{8\pi^2}{2l+1}  \delta_{ll'} \delta_{mm'} \delta_{nn'}
\end{equation}
which immediately gives us
\begin{equation}
       K_{\mu lm; \nu l'm'} = \delta_{ll'} \delta_{mm'} K_{\mu \nu}(l).
\end{equation}


\section{Gauge Invariance and Current Conservation}
\label{app:cons}

The response of the system must be invariant under a gauge
transformation of the perturbing electromagnetic field.  Given an
arbitrary function $\chi(\vO,t)$, a gauge transformation is
given by
\begin{eqnarray}
  \Phi \longrightarrow \Phi - \frac{\partial \chi}{\partial t} \\
  \vec A \longrightarrow \vec A + \vec \nabla \chi.
\end{eqnarray}
If we expand $\chi$ into its frequency and angular momentum components
\begin{equation}
  \chi(\vO,t) = \frac{1}{2\pi} \int d\omega  e^{- i\omega t} \sum_{lm}
Y_{lm}(\vO) \chi_{lm}(\omega)
\end{equation}
and use the definitions of $\cala$ (Eqs. (\ref{eq:vcaladef}) and
(\ref{eq:cala0def})) as well as Eq.  (\ref{eq:calaspheredef}) and the
definitions of the basis $\TT{}$ (Eqs. (\ref{eq:T1def2}) and
(\ref{eq:T2def2})) we find that the gauge transformation results in a
transformation of the angular vector potential $\cala$ given by
\begin{eqnarray}
  \cala_{0lm} &\longrightarrow& \cala_{0lm} - i \omega \chi_{lm}(\omega) \\
  \cala_{1lm} &\longrightarrow& \cala_{1lm} \\
  \cala_{2lm} &\longrightarrow& \cala_{2lm} - \sqrt{l(l+1)} \,
\chi_{lm}(\omega).
\end{eqnarray}
Since the current $\Gamma_\mu$ must be unchanged under this arbitrary
gauge transformation, we have from Eqs. (\ref{eq:sphereresponse}) and
(\ref{eq:Kdelta})
\begin{equation}
  \label{eq:gaugeinvariance}
  i\omega K_{\mu \, 0}(l,\omega) + \sqrt{l(l+1)}  K_{\mu \,
    2}(l,\omega) = 0.
\end{equation}

On the other hand, current conservation demands that
\begin{equation}
   \vec \nabla \cdot \vec J + \frac{\partial \den}{\partial t} =0
\end{equation}
where the current $\vec J$ is given by
\begin{equation}
  \vec J = - \vO \times \vec \Gamma =  \frac{1}{m_b} \psi_e^+(\vO)
\left[ -i \vec \nabla + e \vec A \right]  \psi_e(\vO).
\end{equation}
Recalling that $\Gamma_0$ is just
the fluctuation in density, the current conservation equation becomes
\begin{equation}
  \vec \nabla \cdot (\vO \times \vec \Gamma) - \frac{\partial
    \Gamma_0}{\partial t} = 0 .
\end{equation}
We can rewrite this in  terms of the angular momentum and frequency components
\begin{equation}
 \label{eq:cons7}
  i \omega \Gamma_{0lm}(\omega) + \sum_j  \vec \nabla \cdot (\vO \times
\TT{jlm}) \Gamma_{jlm}(\omega)  = 0
\end{equation}
where the index $j$ is summed over the values $1$ and $2$.  Using the
definition of the $\TT{}$ basis (Eqs. (\ref{eq:T1def2}) and
(\ref{eq:T2def2})) we can derive\cite{Edmonds,Normand}
\begin{eqnarray}
  \vec \nabla \cdot \TT{1lm} &=& \frac{-\sqrt{l(l+1)}}{r} Y_{lm} \\
  \vec \nabla \cdot \TT{2lm} &=& 0
\end{eqnarray}
which then can be used with the current conservation equation
(\ref{eq:cons7}) to yield
\begin{equation}
  \label{eq:precur}
  i \omega \Gamma_{0lm}(\omega) - \sqrt{l(l+1)} \Gamma_{2lm}(\omega) =0
\end{equation}
and hence
\begin{equation}
  \label{eq:currentconservation}
  i \omega K_{0 \, \nu}(l,\omega) -  \sqrt{l(l+1)} K_{2 \, \nu}(l,\omega) = 0.
\end{equation}

\section{The Retarded Correlation Function}
\label{app:ret}

Here we calculate the retarded correlation function for a system of
$N$ noninteracting quasiparticles of mass $m_b$ on a sphere around a
monopole of magnetic charge $2\Delta S$ flux quanta.  It should be
noted that in the calculation of the na\"{\i}ve response $\tilde
\sigma^{\mbox{\small n}}= [\tilde \rho^{\mbox{\small n}}]^{-1}$, we
must replace $m_b$ by $m^*$ everywhere in this calculation.  Since the
eigenstates of this system are given by the monopole spherical
harmonics of Wu and Yang\cite{WuYang} (see Sec. \ref{sub:MFT}) we can
expand the quasiparticle operator in this set of
eigenstates\cite{ChenFetter}
\begin{equation}
  \psi^+(\vO,t) = \sum_{lm} [Y^{(\Delta S)}_{lm}(\vO)]^* c^+_{lm}(t)
\end{equation}
where $c^+_{lm}$ creates a quasiparticle in the $l,m$ state, and the
sum is over all quasiparticle eigenstates.
We  also define
\begin{equation}
   \hatn  \Gamma^{\Delta \vcala}= \frac{1}{m_b}  \left[\vhcalp +
e \Delta \vcala \right]
\end{equation}
such that
\begin{equation}
  \vec \Gamma^{\Delta \vcala}(\vO)  =   \psi^+(\vO) \hatn \Gamma^{\Delta
\vcala} \psi(\vO).
\end{equation}
In addition, we define
\begin{equation}
    \hat \Gamma^{\Delta \vcala}_0 = 1
\end{equation}
for this calculation which yields
\begin{equation}
    \Gamma_0^{\Delta \vcala} = \psi^+(\vO) \psi(\vO) = \den(\vO)
\end{equation}
which differs from the proper definition (\ref{eq:Gamma0def}) by a
constant.  This difference can be neglected here since the commutator (Eq.
\ref{eq:D0def}) of a constant with anything is zero.

With these
definitions, we can now write the retarded correlation function
(\ref{eq:D0def}) as
\begin{eqnarray}
  \nonumber D^0_{\alpha \beta }&(&\vO,t; \vO',t') = -i \theta( t - t')
  \sum_{\shortstack{$\scriptstyle pqp'q'$ \\$\scriptstyle rsr's'$}}
  \left[ \rule[-2ex]{0ex}{6ex}
  Y_{pr}^{(\Delta S)*}(\vO) \hat \Gamma^{\Delta \vcala}_{\alpha}
  Y_{qs}^{(\Delta S)}(\vO) Y_{p'r' }^{(\Delta S)*}(\vO') \hat
  \Gamma^{\Delta \vcala}_{\beta} Y_{q's'}^{(\Delta S)}(\vO') \right.  \\
  & & \nonumber   \left\langle F \left|
  c^{+}_{pr}(t) c_{qs}(t) c^{+}_{p'r'}(t') c_{q's'}(t')
  \right| F \right\rangle   -
   Y_{pr}^{(\Delta S)*}(\vO') \hat \Gamma^{\Delta \vcala}_{\beta}
   Y_{qs}^{(\Delta S)}(\vO') Y_{p'r' }^{(\Delta S)*}(\vO) \hat
  \Gamma^{\Delta \vcala}_{\alpha} Y_{q's'}^{(\Delta S)}(\vO)  \\
  & & \left.  \left\langle F \left|
  c^{+}_{pr}(t') c_{qs}(t') c^{+}_{p'r'}(t) c_{q's'}(t)
  \right| F \right\rangle \rule[-2ex]{0ex}{6ex} \right]
\end{eqnarray}
where $F$ the full Fermi sea (ie the ground state of quasiparticles
filled up to the Fermi level).  The matrix element is clearly zero
unless $(pr)=(q's')$ are states below the Fermi level, and
$(qs)=(p'r')$ are states above the Fermi level. Thus we have
\begin{eqnarray}
  \nonumber D^0_{\alpha \beta }(\vO,t; \vO',t')&=& -i \theta( t - t')
  \sum_{pr}^{\mbox{below}} \sum_{qs}^{\mbox{above}} \left[
  \rule[-2ex]{0ex}{6ex}  e^{i(t-t')(\omega_p - \omega_q)} \right.
\\ \nonumber  &&
Y_{pr}^{(\Delta S)*}(\vO) \hat \Gamma^{\Delta
    \vcala}_{\alpha} Y_{qs}^{(\Delta S)}(\vO)
    Y_{qs}^{(\Delta
    S)*}(\vO') \hat \Gamma^{\Delta \vcala}_{\beta} Y_{pr}^{(\Delta
    S)}(\vO')    - e^{i(t'-t)(\omega_p - \omega_q)}
\\  &&
  Y_{pr}^{(\Delta S)*}(\vO') \hat \Gamma^{\Delta \vcala}_{\beta}
  Y_{qs}^{(\Delta S)}(\vO')
Y_{qs}^{(\Delta S)*}(\vO) \hat
  \Gamma^{\Delta \vcala}_{\alpha} Y_{pr}^{(\Delta S)}(\vO) \left.
  \rule[-2ex]{0ex}{6ex} \right]
\end{eqnarray}
where we have also used the law for the time propagation of creation
and annihilation operators.  Transforming into angular momentum and
frequency components as defined in Eq. (\ref{eq:Dsphere}) now yields
\begin{equation}
  D^0_{\mu l m, \nu l' m'}(\omega) = \sum_{p}^{\mbox{below}}
  \sum_{q}^{\mbox{above}} \left[ \frac{M_{\mu l m, \nu l'
      m'}(p,q)}{(\omega+i0^+) -(\omega_q - \omega_p) } -
  \frac{M_{\mu l m, \nu l' m'}(q,p)}{(\omega+i0^+) + (\omega_q -
    \omega_p)} \right]
\end{equation}
where $\omega_p$ is given by Eq. (\ref{eq:omsubp}),
\begin{equation}
  \label{eq:Mdef}
   M_{\mu l m, \nu l'
      m'}(p,q) = \sum_{rs} N_{\mu l m }(p,r,q,s) N_{\nu l' m'}^* (p,r,q,s)
\end{equation}
and
\begin{equation}
  \label{eq:Ndef}
  N_{\mu l m}(p,r,q,s) = \int d\vO \, [\TT{\mu l m}(\vO)]^* \cdot  \left[
Y_{pr}^{(\Delta S)*}(\vO) \, \hatn\Gamma^{\Delta
    \vcala}\, Y_{qs}^{(\Delta S)}(\vO) \right]
\end{equation}
where the $\TT{}$ and $\hatn \Gamma$ are four-vectors.  The
evaluation of the $\mu= 0$ element of this vector is quite simple
using the identity derived in Appendix \ref{app:ident}:
\begin{eqnarray}
  \label{eq:ident}
  N_{0 l m}(&p&,r,q,s) = \int d\vO \, Y^*_{lm}(\vO) Y_{pr}^{(\Delta
    S)*}(\vO) Y_{qs}^{(\Delta S)}(\vO)  \\  \label{eq:ident2}
                     &=& \left[  \frac{(2l+1)(2p+1)(2q+1)}{4\pi}
\right]^{\frac{1}{2}} (-1)^{s+\Delta S}
                     \left( \begin{array}{ccc} l & p & q \\
                                               -m & -r & s
                             \end{array} \right)
                      \left( \begin{array}{ccc} l & p & q \\
                                                0  & \Delta S  & -\Delta S
                             \end{array} \right)
\end{eqnarray}
where the braces are Wigner $3j$ symbols\cite{Edmonds,Normand}.  Now in order
to calculate the other elements $N_i$, we consider the natural angular
momentum operator for the monopole system\cite{WuYang,OlsenWu}
\begin{eqnarray}
  \hatn L &=& \vec r \times (\vec p + e \Delta \vec A) - \hatn r
  \Delta S \\ \label{eq:Luse} &=& m_b \hatn \Gamma^{\Delta \vcala} +
  \mbox{radial component}.
\end{eqnarray}
This operator acts on the monopole spherical harmonic $Y_{qs}^{(\Delta
  S)}$ to give the monopole vector harmonics of Olsen, Osland, and
Wu\cite{OlsenWu}:
\begin{equation}
   \hatn L Y_{qs}^{(\Delta S)}(\vO) = \sqrt{q(q+1)} \vec  Y_{qqs}^{(\Delta
S)}(\vO)
\end{equation}
where the monopole vector harmonic is defined as
\begin{equation}
  \label{eq:mvecharm}
   \vec  Y_{jlm}^{(\Delta S)}(\vO) = \sum_{n,\alpha} \langle l 1 n \alpha
  | j  m \rangle Y^{(\Delta S)}_{l n}(\vO)
  \hatn e_{\alpha.}
\end{equation}
Here, the Clebsch-Gordan coefficients are defined as in
Ref.\cite{Normand}, and the spherical tensor vector basis $ \hatn
e_{\alpha}$ is given by
\begin{eqnarray}
  \hatn e_{\pm 1} &=& \mp \frac{1}{\sqrt{2}} (\hatn e_x \pm i \hatn e_y) \\
  \hatn e_0      &=& \hatn e_z
\end{eqnarray}
where $\{ \hatn e_x,\hatn e_y,\hatn e_z \}$ are the usual
three-dimensional Cartesian basis vectors.  In the case of $\Delta S
=0$, Eq. (\ref{eq:mvecharm}) defines the usual\cite{Edmonds,Normand} spherical
vector
harmonics $\vec Y_{jlm} \equiv \vec Y_{jlm}^{(0)}$ in terms of the
usual spherical harmonics $Y_{l n} \equiv Y^{(0)}_{l n}$.

Using the relation (\ref{eq:Luse}) we now have
\begin{eqnarray}  \nonumber
  Y_{pr}^{(\Delta S)*}(\vO) \, \hatn \Gamma^{\Delta \vcala}\,
  Y_{qs}^{(\Delta S)}(\vO) &=& \frac{1}{2m_b} \left[ \sqrt{q(q+1)}
  Y_{pr}^{(\Delta S)*}(\vO) \vec Y_{qqs}^{(\Delta S)}(\vO) \right. \\  & & +
  \left. \sqrt{p(p+1)} \vec Y_{ppr}^{(\Delta S)*}(\vO) Y_{qs}^{(\Delta
    S)}(\vO) \right] + \mbox{radial component}
\end{eqnarray}
where we have left-right symmetrized the operator.  Since the unknown
radial component can not contribute to the response of a system on the
sphere, we can safely drop this piece (as long as we don't try to
calculate a radial ($\mu = 3$) response.

We now
note that the $\TT{}$ functional vector basis can be written in terms
of the usual spherical vector harmonics as\cite{Normand}
\begin{equation}  \label{eq:Tdef}
  \TT{ilm} =\sum_k  C_{ik} \vec Y_{l (l+k) m}
\end{equation}
where $i=1,2,3$, the index  $k$ is summed over the values $-1,0,1$, and
\begin{eqnarray}
     \nonumber
   k&  = & \; \;\; \;  -1 \;\;\; \; \;\;\; 0 \;\;\;\;\;\;\;  -1   \\
 \label{eq:Cdef}  C_{ik} & = &    \left(
\begin{array}{ccc} \sqrt{\frac{l+1}{2l+1}} & 0 & \sqrt{\frac{l}{2l+1}} \\
             0 & -i & 0 \\
                  \sqrt{\frac{l}{2l+1}} & 0 & -\sqrt{\frac{l+1}{2l+1}}
                  \end{array}  \right)_.
\end{eqnarray}
Now using this relation, the definition (\ref{eq:mvecharm}) of the
monopole vector harmonics, and the same identity (\ref{eq:ident2})
derived in Appendix \ref{app:ident}, then converting Clebsch-Gordan
coefficients to $3j$ coefficients\cite{Edmonds,Normand}, we find that  Eq.
(\ref{eq:Ndef}) can be rewritten for
$i=1,2$ as
\begin{equation}
  N_{i  l m}(p,r,q,s) =  \sum_k C_{i k} Q_{l,l+k,m}(p,r,q,s)
\end{equation}
where $k$ is summed over the values  $-1,0,1$, and
\begin{eqnarray}   \nonumber
    Q_{l,j,m}(p,r,q,s)&&  = \frac{1}{2m_b} \left[\frac{
(2l+1)(2j+1)(2p+1)(2q+1)}
{4 \pi} \right]^{\frac{1}{2}} \left( \begin{array}{ccc} j  & p  & q \\
                                                 0  &  \Delta S  & -\Delta S
                             \end{array} \right) \\ \nonumber
                             \left( \begin{array}{ccc} l  & p & q \\
                                                      -m   &  -r & s
                             \end{array} \right)
                           &&  \left[  \sqrt{p(p+1)(2p+1)}
                               (-1)^{(p+q+j+s+\Delta S)}
                         \left\{ \begin{array}{ccc} l  & p & q \\
                                                    p   & j & 1
                             \end{array} \right\}
                          \right. \\ \label{eq:Qljm}
                  &&  + \left. \sqrt{q(q+1)(2q+1)}
                               (-1)^{(3q +1 + \Delta S -s + l + p)}
            \left\{ \begin{array}{ccc} l  & q & p \\
                                                    q   & j  & 1
                             \end{array} \right\}
 \right]_.
\end{eqnarray}
The brackets here denote Wigner $6j$ symbols, and we have used the
orthogonality identity\cite{Edmonds}
\begin{eqnarray} \nonumber
 \sum_{n_1, n_2, n_3} (&-1&)^{(l_1+l_2+l_3+n_1+n_2+n_3)}
 \left( \begin{array}{ccc} j_1  & l_2 & l_3 \\
                             m_1  &  n_2 & -n_3
                             \end{array} \right)
 \left( \begin{array}{ccc} l_1  & j_2 & l_3 \\
                             -n_1  &  m_2 & n_3
                             \end{array} \right)
 \left( \begin{array}{ccc} l_1  & l_2 & j_3 \\
                             n_1  &  -n_2 & m_3
                             \end{array} \right)  \\
&=&
   \left( \begin{array}{ccc} j_1  & j_2 & j_3 \\
                             m_1  &  m_2 & m_3
                             \end{array} \right)
 \left\{ \begin{array}{ccc} j_1  & j_2 & j_3 \\
                            l_1 & l_2 & l_3
                             \end{array} \right\}
\end{eqnarray}
and the symmetry properties of the $3j$ symbols\cite{Edmonds,Normand}
to derive Eq. (\ref{eq:Qljm}).  We now note that each $N_{\mu}$ contains a
Wigner $3j$ coefficient with indices $r$ and $s$ that are then summed
over in Eq. (\ref{eq:Mdef}).  Using the orthogonality
relation\cite{Edmonds,Normand}
\begin{equation}
  \sum_{r,s} \left( \begin{array}{ccc} l  & p  & q \\
                                                   -m   & -r  & s
                             \end{array} \right)
\left( \begin{array}{ccc} l'  & p  & q \\
                                                   -m'   & -r  & s
                             \end{array} \right) = \frac{\delta_{ll'}
\delta_{mm'}}{2l+1}
\end{equation}
we perform the sum in  (\ref{eq:Mdef}) such that
\begin{equation}
   M_{\mu l m, \nu l' m'}(p,q) = M_{\mu,\nu}(l,p,q) \delta_{ll'}
  \delta_{mm'}
\end{equation}
and
\begin{equation}
\label{eq:xx1}
M_{\mu,\nu}(l,p,q) = \left[\frac{(2p+1)(2q+1)}{4\pi} \right] \tilde
N_{\mu}(p,q) \tilde N_{\nu}^*(p,q)
\end{equation}
where
\begin{eqnarray}
\label{eq:tildeNdef}
  \tilde N_0(p,q) &=&  \left( \begin{array}{ccc} l  & p  & q \\
                                                0   & \Delta S  & -\Delta S
                             \end{array} \right)  \\
  \tilde N_i(p,q) &=& \sum_k C_{ik} \tilde  Q_{l,l+k}(p,q).
\end{eqnarray}
Once again, $k$ is summed over $-1,0,1$, the matrix $C$ is the same conversion
matrix from Eq (\ref{eq:Cdef}), $i$
takes on the values $1,2$, and now
\begin{eqnarray}
  \tilde Q_{l,j}(p,q) &=& \nonumber \frac{1}{2m_b} \sqrt{2j+1}
  \left( \begin{array}{ccc} j  & p  & q \\
                          0   & \Delta S  & -\Delta S
                             \end{array} \right)
   \left[ (-1)^{p+q+j} \sqrt{p(p+1)(2p+1)}
 \left\{ \begin{array}{ccc} l  & p  & q \\
                            p  & j  &  1
                             \end{array} \right\} \right.
                    \\   & & \,\,\,   + \left.
                       (-1)^{q-p+l+2\Delta S +1} \sqrt{q(q+1)(2q+1)}
 \left\{ \begin{array}{ccc} l  & q  & p \\
                            q  & j  &  1
                             \end{array} \right\} \right]_.
\end{eqnarray}
Finally, collecting our result, we have
\begin{equation}
\label{eq:xxthis}
  D^0_{\mu,\nu}(l,\omega) = \sum_{p}^{\mbox{below}}
  \sum_{q}^{\mbox{above}} \left[ \frac{M_{\mu \nu}(l,p,q)}
{(\omega+i0^+) -(\omega_q - \omega_p) } -
  \frac{M_{\mu \nu}(l,q,p)}{(\omega+i0^+) + (\omega_q -
    \omega_p)} \right]
\end{equation}
where the $p$ sum is over states such that $\omega_p$ is less than or
equal to the Fermi energy, and the $q$ sum is over states such that
$\omega_q$ is greater than the Fermi energy.


\section{Coupling of Three Monopole Harmonics}
\label{app:ident}

The monopole spherical harmonics $Y^{(q)}_{lm}(\theta,\phi)$ of Wu and
Yang\cite{WuYang} can be written in terms of the rotation matrices
$D^l_{m'm}(\phi,\theta,\psi)$ of ordinary quantum
mechanics\cite{Normand}.  The relation between the two is given by
\begin{equation}
  Y^{(q)}_{lm}(\theta,\phi) = \left[ \frac{2l+1}{4\pi} \right]^{\frac{1}{2}}
[D^l_{m,-q}(\phi,\theta,-\phi)]^*_.
\end{equation}
Now, using the coupling relation for three rotation
matrices\cite{Edmonds,Normand}
\begin{eqnarray} \nonumber
  \frac{1}{8 \pi^2} \int_0^{2\pi}  d\phi \int_{-1}^{1} d\cos \theta
&&\int_0^{2\pi} d\psi \,  D^{j_1}_{m_1 m_1'}(\phi,\theta,\psi) D^{j_2}_{m_2
m_2'}(\phi,\theta,\psi) D^{j_3}_{m_3 m_3'}(\phi,\theta,\psi)    \\  &&=
  \left( \begin{array}{ccc} j_1  & j_2  & j_3 \\
                             m_1   & m_2  & m_3 \end{array} \right)
                          \left( \begin{array}{ccc}  j_1  & j_2  & j_3  \\
                            m_1'   & m_2'  & m_3'
                             \end{array} \right)
\end{eqnarray}
and the decomposition of the rotation matrix,
\begin{equation}
  D^l_{m'm}(\phi,\theta,\psi) = e^{-i\phi m'} d^j_{m'm}(\theta) e^{-i\psi m}
\end{equation}
we can easily derive the corresponding law for monopole harmonics
\begin{eqnarray} \nonumber
  \int_0^{2 \pi} d\phi \int_{-1}^{1} &&  d \cos \theta  \,
Y^{(q_1)}_{l_1m_1}(\theta,\phi)
 Y^{(q_2)}_{l_2m_2}(\theta,\phi) Y^{(q_3)}_{l_3m_3}(\theta,\phi) =  \\
&& \left[ \frac{(2l_1+1)(2l_2+1)(2l_3+1)}{4\pi} \right]^{\frac{1}{2}}
  \left( \begin{array}{ccc} l_1  & l_2  & l_3 \\
                             -m_1   & -m_2  & -m_3 \end{array} \right)
                          \left( \begin{array}{ccc}  l_1  & l_2  & l_3  \\
                            q_1   & q_2  & q_3
                             \end{array} \right)_.
\end{eqnarray}
Finally, we use the relation\cite{WuYang}
\begin{equation}
   [Y^{(q)}_{lm}(\theta,\phi)]^* = (-1)^{q+m} Y^{(-q)}_{l,-m}(\theta,\phi)
\end{equation}
and the fact that $Y^{(0)}_{lm} \equiv Y_{lm}$ is just the usual
spherical harmonic to trivially derive Eq. (\ref{eq:ident2}).

\section{Interaction Coefficients}
\label{app:VW}

An arbitrary rotationally symmetric function $v(|\vO-\vO'|)$ on the unit
sphere can be expanded in terms of spherical harmonics
as\cite{Edmonds,Normand}
\begin{equation}
  \label{eq:vvv}
  v(\vO,\vO') = 4 \pi \sum_{lm} Y^*_{lm}(\vO) Y_{lm}(\vO') f_l
\end{equation}
where
\begin{equation}
   f_l = \frac{1}{2}\int_{-1}^{1} \,  d\!\cos \theta \,\,  P_l(\cos \theta) \,
v \! \left( [2-2\cos \theta]^{\frac{1}{2}} \right)_.
\end{equation}
and $P_l$ is the Legendre polynomial\cite{Grad}.  For the Coulomb
interaction $v=1/|\vO - \vO'|$, for example, it is a well known result
that\cite{Edmonds}
\begin{equation}
\label{eq:vl}
  v(l) \equiv 4 \pi  f_l^{\mbox{\small Coulomb}}  = \frac{4\pi}{2l+1.}
\end{equation}

In order to calculate the Chern-Simons interaction coefficient, we begin by
considering the function
\begin{equation}
  \label{eq:flldef}
\frac{1}{2} \ln \left( |\vO - \vO'|^2 \right) = 2\pi \sum_{lm}  Y^*_{lm}(\vO)
Y_{lm}(\vO') f_l
\end{equation}
where
\begin{equation}
  f_{l} = \frac{1}{2} \int_{-1}^{1} dz \,  P_l(z) \,  \left[ \ln 2 +
\ln(1-z)\right].
\end{equation}
Now, since $P_0$ is a constant, and the Legendre polynomials form an
orthogonal set\cite{Grad}, the first term vanishes except when $l=0$,
which is a case that we will not be concerned with. To evaluate this
integral, we first change the integration variable to $-z$, using the
fact that $P_l(-z) = (-1)^l P_l(z)$. Next, we  use the well known identity
\begin{equation}
  \ln x = \lim_{\alpha \rightarrow 0} \frac{x^\alpha -1}{\alpha}
\end{equation}
to rewrite the integral  as the following limit for $l \ne 0$,
\begin{equation}
  f_{l} = \frac{(-1)^l}{2} \int_{-1}^{1} dz \, P_l(z) \, \lim_{\alpha
  \rightarrow 0} \left[ \frac{(1+z)^{\alpha} -1}{\alpha} \right]_.
\end{equation}
 The
second term is a constant with respect to $z$ and, as discussed above,
integrates to zero except in the $l=0$ term.  The first term is an
integral that can be found in a standard table\cite{Grad} to yield
\begin{equation}
    f_l = (-1)^l \lim_{\alpha \rightarrow 0} \left[ \frac{2^{\alpha + 1} \left[
\Gamma(\alpha+1) \right]^2}{\alpha \Gamma(\alpha+l+2) \Gamma(\alpha+1-l)}
\right]_.
\end{equation}
The limit can now be taken by using the reflection principle  that
$\Gamma(z) \Gamma(1-z) \sin (\pi z) = \pi$ to give
\begin{equation}
   f_l = \frac{1}{l(l+1)}
\end{equation}
for $l \ne 0$.  Using this result along with Eq. (\ref{eq:flldef}), we
then derive the useful identity
\begin{equation}
  \frac{1}{2} \nabla_{\vO} \ln \left( | \vO - \vO'|^2 \right) = \frac{
    \vO - \vO'} {|\vO - \vO'|^2} = 2 \pi \sum_{lm} \TT{1lm}^*
  Y_{lm}(\vO') \frac{1}{\sqrt{l(l+1)}}
\end{equation}
where we have used Eq. (\ref{eq:T1def2}) to take the gradient of the
spherical harmonic.

We are now interested in the angular momentum components of the quantity
\begin{equation}
  \vec W(\vO,\vO') = - \frac{\vO \times (\vO \times \vO')} {|\vO -
    \vO'|^2} = \frac{(\vO' - \vO)}{|\vO - \vO'|^2} + \frac{\vO}{2}
\end{equation}
where we have used various simple vector identities\cite{Grad} to
obtain this form. Since the second term is purely radial, it will only
contribute to the $\TT{3}$ component which we have already completely
decoupled.  Thus, we can safely drop this piece.  So, converting into
angular momentum components yields for $i=1,2$, and $l \ne 0$,
\begin{eqnarray}
  W_{ilm;0l'm'} &=& \int d\vO \int d\vO' \TT{ilm}(\vO) \cdot \vec
  W(\vO,\vO') Y^*_{lm}(\vO') \\ &=& \int d\vO \int d\vO' \TT{ilm}(\vO)
  \cdot \left[\sum_{kn} \TT{1kn}^*
  Y_{kn}(\vO')\frac{-2\pi}{\sqrt{k(k+1)}} \right] Y^*_{l'm'}(\vO') \\
  \label{eq:Wres}
  &=& - \delta_{1i} \delta_{ll'} \delta_{mm'} \frac{2\pi}{\sqrt{l(l+1)}.}
\end{eqnarray}

\section{Effective Electric and Magnetic Fields}
\label{app:EB}

Here we write the induced vector potential in terms of  its angular momentum
components as described in Eqs. (\ref{eq:Ain}), (\ref{eq:Arec0}), and
(\ref{eq:Arec}):
\begin{eqnarray}
   A_0^{\mbox{\small in}}(\vO)_{\mbox{\small Coulomb}} &=& {\scriptstyle \left(
\frac{N-1}{N} \right)} \sum_{lm} Y_{lm}(\vO) v(l) \Gamma_{0lm} \\
   A_0^{\mbox{\small in}}(\vO)_{\mbox{\small Chern-Simons}} &=& {\scriptstyle
\left( \frac{N-1}{N} \right)} \sum_{lm} Y_{lm}(\vO) w(l) \Gamma_{1lm} \\
   \vec A^{\mbox{\small in}}(\vO) &=& {\scriptstyle \left( \frac{N-1}{N}
\right)} \sum_{lm} \TT{2lm}(\vO) w(l) \Gamma_{0lm}
\end{eqnarray}
where here $\Gamma_{\mu l m}$ represents the expectation $\langle
\Gamma_{\mu l m} \rangle$ and we have used the definition of the
$\TT{}$ basis.  We will now establish that associated electric and
magnetic fields are those given by Eqs. (\ref{eq:CSe}), (\ref{eq:CSb})
and (\ref{eq:Coulin})

We calculate the Chern-Simons magnetic field
\begin{eqnarray}
  \vec b^{\mbox{\small in}}(\vO) &=&\vO \cdot  [\vec \nabla \times \vec
A^{\mbox{\small in}}(\vO)] \\
  &=& {\scriptstyle \left( \frac{N-1}{N} \right)} \sum_{lm} w(l) \Gamma_{0lm}
\vO \cdot [\vec \nabla \times \TT{2lm}(\vO)].
\end{eqnarray}
Using a vector identity\cite{Normand} of the $\TT{}$ basis this becomes
\begin{eqnarray}
 \vec b^{\mbox{\small in}}(\vO) &=& {\scriptstyle \left( \frac{N-1}{N} \right)}
\sum_{lm} \tilde \phi  \frac{2\pi}{\sqrt{l(l+1)}} \Gamma_{0lm} \sqrt{l(l+1)}
Y_{lm}(\vO) \vO\\
  &=&  {\scriptstyle \left( \frac{N-1}{N} \right)} \, 2 \pi \tilde \phi \,
\left\langle  \Gamma_0(\vO)  \right\rangle \vO
\end{eqnarray}
which agrees with Eq. (\ref{eq:CSb}).  Similarly we can calculate the
Chern-Simons electric field
\begin{eqnarray}
  \vec e^{\mbox{\small in}}(\vO) &=& \vec \nabla A_{0}^{\mbox{\small
      in}}(\vO)_{\mbox{\small Chern-Simons}} - \frac{\partial}{\partial t} \vec
A^{\mbox{\small in}}(\vO) \\
  &=& {\scriptstyle \left( \frac{N-1}{N} \right)} \sum_{lm} w(l) \left[ \vec
\nabla Y_{lm}(\vO) \Gamma_{1lm} - \TT{2lm}(\vO) \frac{\partial}{\partial t}
\Gamma_{0lm} \right].
\end{eqnarray}
Using the definition (\ref{eq:T1def2}) of the $\TT{}$ basis and the current
conservation equation (\ref{eq:precur}) this can be rewritten as
\begin{eqnarray}
  \vec e^{\mbox{\small in}}(\vO) &=& {\scriptstyle \left( \frac{N-1}{N}
\right)} \sum_{lm} \tilde \phi  \frac{2\pi}{\sqrt{l(l+1)}} \left[\TT{1lm}(\vO)
\Gamma_{1lm} + \TT{2lm}(\vO) \Gamma_{2lm} \right]  \sqrt{l(l+1)} \\
            &=& {\scriptstyle \left( \frac{N-1}{N} \right)} \, 2 \pi \tilde
\phi \, \left\langle \vec \Gamma(\vO) \right\rangle
\end{eqnarray}
in accordance with Eq. (\ref{eq:CSe}).  Finally we have the Coulombic
electric field
\begin{equation}
\label{eq:coulfinal}
  \vec E^{\mbox{\small in}}(\vO) = {\scriptstyle \left( \frac{N-1}{N} \right)}
\sum_{lm} \vec \nabla Y_{lm}(\vO) v(l) \Gamma_{0lm}.
\end{equation}
We can not take the divergence of this expression directly to verify
Eq. (\ref{eq:Coulin}) since there is some component of the electric
field which is normal to the surface of the sphere that is not included in
(\ref{eq:coulfinal}) (we don't care about this component since it does not
couple to our problem).  However, it is a trivial application of
electrodynamics to use Eqns. (\ref{eq:vvv}), (\ref{eq:vl}), and
(\ref{eq:intCoul}) to establish that this electric field does indeed
correspond to the charge density $\langle \Gamma_0 \rangle$ and thus
satisfies Eq.  (\ref{eq:Coulin}).



\begin{references}
\bibitem{end1} In Ref. \cite{Simon} we were also concerned with
satisfying Kohn's theorem --- another result of Galilean invariance
--- in the planar geometry.  As far as we know, there is no simple
analog to Kohn's theorem in the spherical geometry.
\end{references}
\end{document}